\documentclass[envcountsame,runningheads]{llncs}
\usepackage{amsmath}
\usepackage{amssymb}
\DeclareFontFamily{U}{mathb}{\hyphenchar\font45}
\DeclareFontShape{U}{mathb}{m}{n}{
      <5> <6> <7> <8> <9> <10> gen * mathb
      <10.95> mathb10 <12> <14.4> <17.28> <20.74> <24.88> mathb12
      }{}
\DeclareSymbolFont{mathb}{U}{mathb}{m}{n}
\DeclareMathSymbol{\precneq}{3}{mathb}{"AC}
\newcommand{\reduceq}{\preceq}
\newcommand{\reducneq}{\precneq}
\newcommand{\notreduceq}{\npreceq}
\newcommand{\mysmall}[1]{\raisebox{0.2ex}{$\scriptscriptstyle#1$}}
\newcommand{\IR}{\mathbb{R}}
\newcommand{\IP}{\mathbb{P}}
\newcommand{\IC}{\mathbb{C}}
\newcommand{\IA}{\mathbb{A}}
\newcommand{\IT}{\mathbb{T}}
\newcommand{\IQ}{\mathbb{Q}}

\newcommand{\IF}{\mathbb{F}}
\newcommand{\Q}{\mathbb{Q}}
\newcommand{\IZ}{\mathbb{Z}}
\newcommand{\IH}{\mathbb{H}_{\IR}}
\newcommand{\IN}{\mathbb{N}}
\newcommand{\dom}{\operatorname{dom}}
\newcommand{\lcm}{\operatorname{lcm}}
\newcommand{\person}[1]{\textsc{#1}}
\newcommand{\COMMENTED}[1]{}

\newcommand{\mycite}[2]{\cite[\textsc{#1}]{#2}}
\newcommand{\BCSS}{BSS}
\newtheorem{fact}[theorem]{Fact}

\newtheorem{scholiumf}[theorem]{Scholium\footnotemark}
\newtheorem{myproblem}{Problem}
\newcommand{\SQ}{\mathbb{S}\IQ}
\newcommand{\QP}{\IQ_{P}}
\newcommand{\NP}{\ensuremath{\mathcal{NP}}}
\def\ri{\IR^{\infty}}

\begin{document}
\title{An Explicit Solution to \\ \person{Post}'s Problem over the Reals}
\titlerunning{An Explicit Solution to Post's Problem over the Reals}
\authorrunning{K.~Meer and M.~Ziegler}
\tocauthor{K.~Meer and M.~Ziegler}
\author{Klaus Meer\inst{1}\thanks{partially supported 
by  the IST Programme of the European
Community, under the PASCAL Network of Excellence, IST-2002-506778
and by the Danish Natural Science Research Council SNF.
This publication only reflects the authors' views.}
and Martin Ziegler\inst{1,2}\thanks{%
M. Ziegler's stay in Odense was made possible by 
project 21-04-0303 of Statens Natur\-videnskabelige Forskningsr{\aa}d SNF;
he is now supported by DFG project \texttt{Zi1009/1-1}.}}
\institute{Department of Mathematics and Computer Science\\
Syddansk Universitet,
Campusvej 55, 5230 Odense M, Denmark \and
PaSCo and HNI, University of Paderborn, 33095 Paderborn, GERMANY.
\email{\{meer,ziegler\}@imada.sdu.dk}
}

\date{}
\maketitle
\begin{abstract}
In the \BCSS{} model of real number computations
we prove a concrete and explicit semi-decidable 
language to be undecidable
yet not reducible from (and thus strictly easier than)
the real Halting Language.
This solution to \person{Post}'s Problem over the reals
significantly differs from its classical, discrete variant
where advanced diagonalization techniques are only known to
yield the \emph{existence} of such intermediate Turing degrees.

Strengthening the above result, we construct 
(that is, obtain again explicitly)
as well an uncountable number
of incomparable semi-decidable Turing degrees below the real Halting problem
in the \BCSS{} model.
Finally we show the same to hold for the linear \BCSS{} model,
that is over $(\IR,+,-,<)$ rather than $(\IR,+,-,\times,\div,<)$.
\end{abstract}
\section{Introduction}
Is every super-Turing computer capable of solving the 
discrete Halting Problem $H$?

More formally, does each undecidable, recursively enumerable language $P\subseteq\IN$,
when serving as oracle to some appropriate Turing Machine $M$,
enable this $M^P$ to decide $H$?
That question of \person{E.L.~Post} from 1944 was answered to
the negative in 1956/57 independently
by \person{Muchnik} and \person{Friedberg} \cite{Friedberg}%
\footnote{The existence of intermediate Turing degrees that
need not to be r.e. follows from a result by Kleene and Post from 1954, 
see \cite{Soare}.}.
Devising the \emph{finite injury priority} sophistication of
diagonalization, they proved the existence of r.e. Turing degrees 
strictly between those of $\emptyset$ and $\emptyset'=H$;
cf. \cite[\textsc{Chapters~V} to \textsc{VII}]{Soare}.

\medskip
While the diagonal language is also based on a mere existence proof,
its reduction to $H$ reveals this as well as many other explicit and
practical problems in automatized software verification undecidable.
In contrast, problems like $P$ are until nowadays only known to exist
but have resisted any explicit, not to mention intuitive, description
--- which is a pity as they can have significant impact to the raising
field of hypercomputation, that is, (theory) of super-Turing computation.
Namely whereas, in spite of e.g. \cite{Yao},
many scientists deny the Halting Problem $H$ to be solvable even by
a \emph{non}-Turing device like \cite{Hogarth2,KieuTCS},
they might be less reluctant towards the solvability of
a problem like $P$ because it is strictly easier than $H$.
However, attempts to actually devise a physical system solving $P$
are futile as long as $P$ itself is known no more than to just exist.

\medskip
It turns out that for real number problems the
situation is quite different. More precisely, for the
$\IR$-machine model due to \person{Blum}, 
\person{Shub}, and \person{Smale} \cite{BSS,BCSS}, we
\emph{explicitly} present a semi-decidable language
(specifically, the set $\IQ$ of rationals) and
prove it to neither be reducible from the
real Halting Problem $\IH$ nor from
the set $\IA$ of algebraic reals.
The proof exploits that real computability theory,
apart from logic as in the discrete case, has
also algebraic and topological aspects.

\medskip
Section~\ref{secBCSS} recalls the basics
of real number computation in the \BCSS{} model as well as the
recursion-theoretic notions
of reducibility and degrees; 
Section~\ref{secSolve} contains the first 
main result of our work; we show 
~$\IQ\reducneq\IA$,
i.e. the real algebraic numbers cannot be decided
using a \BCSS{} oracle machine which has access to the
(undecidable!) set of rationals as oracle set.
Section~\ref{secPos} proves the
`$\reduceq$'-part, Section~\ref{secNeg} the
`$\reducneq$'-part. 
In Section~\ref{secExistence}
the results are generalized in order to get
an uncountable number of incomparable semi-decidable problems
below the real Halting problem.
We conclude in Section~\ref{secConclusion}
with some general remarks on hypercomputation.
\subsection{Related Work}
Our contribution adds to other results, indicating that many
(separation-) problems which seem to require non-constructive
(e.g., diagonalization) techniques in the discrete case,
admit an explicit solution over the reals. For instance,
a problem neither in $\cal{VP}$ nor $\cal{VNP}$-complete
(provided that $\cal{VP}\not=\cal{VNP}$, of course) was presented
explicitly in \mycite{Section~5.5}{ACT2}.

\person{Cucker}'s work \cite{Cucker} is about
the Arithmetic Hierarchy over $\IR$, that is,
degrees beyond the real Halting Problem $\IH$.

\person{Hamkins} and \person{Lewis}
considered \person{Post}'s Problem over the reals for
\emph{Infinite Time} Turing Machines, that is, with respect
to arguments $x\in\IR$ given by their binary expansion
and for hypercomputers performing an ordinal number of steps like
$1,2,3,\ldots,n,\ldots,\omega,\omega+1,\ldots,2\omega,$\ldots
They showed in \cite{Hamkins2} that in this model,
\\ \textbullet ~
   for {sets} of reals the answer is ``no''
   just like in the classical discrete case.
\\ \textbullet ~
   for \emph{single} real numbers $x$ on the other hand,
   considered as sets $L_x\subseteq\IN$ of those indices
   where the binary expansion of $x$ has a $1$,
   there is {no} undecidable degree below
   that of the Halting Problem (of Infinite Time Machines).
   \person{Post}'s Problem therefore is to be answered
   to the \emph{positive} in this latter setting!

The existence of different \emph{complexity degrees} 
below $\NP$ in the \BCSS{} model 
both for real and for complex numbers was 
studied in a series of papers \cite{Ben-David,Chapuis,Malajovich}
and related to classical results (cf. \cite{Ladner,Schoening}) for the Turing model.

\subsection{The \BCSS{} Model of Real Number Computation} \label{secBCSS}

This section summarizes very briefly the main ideas
of real number computability theory. For a more detailed
presentation see \cite{BCSS}.

Essentially a (real) \BCSS{} machine can be considered as a Random Access Machine
over $\IR$ which is able to perform the basic arithmetic operations  at
unit cost and which registers can hold arbitrary real numbers.

\begin{definition}[\cite{BSS}]  \label{defBCSS}
\begin{itemize}
\item[a)] 
Let $ Y \subseteq \ri := \bigoplus_{k \in \IN} \IR^k $, i.e. the set of
finite sequences of real numbers. A {\sc \BCSS{} machine $M$
over $\IR$ with admissible input set $Y$} is given by a finite set
$I$ of
instructions labelled by $ 1,\ldots,N . $ A configuration of $M$
is a quadruple $(n,i,j,x) \in I \times \IN \times \IN \times \ri \ . $ Here, 
$n$ denotes the currently executed instruction, $i$ and $j$ are used as
addresses (copy-registers) and $x$ is the actual content of the registers
of $M$. The initial configuration of $M'$s computation on input $y \in Y$
is $ (1,1,1,y) $ . If $ n = N $ and the actual configuration is
$(N,i,j,x) $, the computation stops with output $x$ . \newline
The instructions $M$ is allowed to perform are of the following types :

\begin{description}
\item[computation:\ ] $n: x_s\leftarrow x_k \circ_n x_l$,
where $\circ_n \in \{+,-,*,:\}$ or \newline
$n: x_s \leftarrow \alpha$ for some constant $ \alpha \in \IR \ . $\newline
The register $x_s$ will get the value $ x_k \circ_n x_l $ or $ \alpha$, 
respectively. 
All other register-entries remain unchanged.
The next instruction will be $n+1$;
moreover, 
the copy-register $i$ is either incremented by one, replaced by $0$, or
remains unchanged. The same holds for copy-register $j$.
\item[branch:\ ] $n$: {\bf if $x_0\geq 0$ goto $\beta(n)$ else goto $n+1$.}
According to the answer of the test the next instruction is determined
(where $ \beta(n) \in I ).$ All
other registers are not changed.
\item[copy:\ ] $n: x_i\leftarrow x_j$, i.e. the content of
the ``read"-register is copied into the ``write"-register. The
next instruction is $n+1$; all other registers remain unchanged.

\end{description}
\item[b)] The size of an $ x \in \IR^k$ 
is $size_{\IR}(x)=k$. The cost of any of the
above operations is $1$. The cost of a computation
is the number of operations performed until the machine halts.

\item[c)]

A set $A \subseteq \ri$ is called a decision problem or a
language over $\ri$.

We call a function $f : A \to \ri$ 
(\BCSS--) computable iff it is realized by a \BCSS{} machine over admissible
input set $A$. Similarly, a set $A \subseteq \ri$ is 
decidable in $\ri$ iff its characteristic function is computable. 
It is semi-decidable iff there is a \BCSS{} algorithm which takes inputs from $\ri$ and
halts precisely on the elements belonging to $A.$

\item[d)] A \BCSS{} oracle machine using an oracle set $B \subseteq \ri$
is a \BCSS{} machine with an additional type of node called oracle node.
Entering  such a node the machine can ask the oracle whether a previously
computed element $x\in\ri$ belongs to $B.$ The oracle gives the correct
answer at unit cost. 
\end{itemize}

\end{definition}
Several further concepts and notions now can be 
defined straightforwardly.

\begin{definition}\label{Definition:Halting-Problem}
The \emph{real Halting Problem} $\IH$ 
is the following decision problem. Gi\-ven 
the code $c_M \in \ri$ of a \BCSS{} machine $M$ together with
an $x \in \ri$, does
$M$ ter\-mi\-nate its computation on input $x?$
\end{definition}
Both the existence of such a coding for \BCSS{} machines and 
the undecidability of $\IH$ in the \BCSS{} model were shown in \cite{BSS}.

Next, oracle reductions are defined as usual.
\begin{definition}\label{Definition:Orakelreduktion}
\begin{itemize}
\item[a)] A real number decision problem $A$ is \emph{reducible}
to another decision problem $B$ if there is a \BCSS{} oracle machine 
that decides membership in $A$ by using $B$ as oracle set.

We denote this reducibility by $A \reduceq B.$ We write
$A \reducneq B$ when $A$ is reducible to $B,$ but $B$ is not reducible
to $A.$
\item[b)] If $A$ is reducible to $B$ and vice versa, we write
$A\equiv B$. This defines equivalence classes 
$\{B: A\equiv B\}$ among real number 
decision problems called 
\emph{(real) Turing degrees} or \emph{\BCSS{} degrees}.

\item[c)]
If none of two problems is reducible to the other,
they are said to be incomparable.
\end{itemize}
\end{definition}
The main question treated in this paper is: Are there
incomparable Turing degrees strictly between 
the degree $\emptyset$ of decidable problems
in $\ri$ and the degree $\emptyset'$ 
of the real Halting problem $\IH?$

\section{Explicit Solution to \person{Post}'s Problem over the Reals} 
\label{secSolve}
Consider the sets $\IQ$ of all rational numbers
and $\IA$ of all algebraic reals, that is,
of real zeros of polynomials with rational coefficients, only.
$\IQ$ is obviously semi-decidable (upon input of $x\in\IR$,
simply check for all pairs of integers $r,s\in\IZ$ whether $x=r/s$)
but well-known  not to be decidable \cite{Herman,Meer}.
In fact the same holds for $\IA$:
Given $x\in\IR$, try for all  polynomials
$p\in\IQ[X]$ whether $p(x)=0$.

Our first main result states that, even given oracle access to $\IQ$,
$\IA$ remains undecidable: $\IA\notreduceq\IQ$.
Since oracle access to the Halting
Problem $\IH$ of \BCSS{} machines allows to decide $\IA$ by
querying whether the above search for $p\in\IQ[X]$ terminates,
$\IQ$ thus constitutes an explicit example of a real \BCSS{} degree
strictly between the decidable one 
and that of the Halting Problem.

We also show ~$\IQ\reduceq\IA$.
\begin{theorem} \label{thMain}
In the \BCSS{} model of real number computation it holds
~$\IQ\reducneq\IA$. In particular, transcendence is not
semi-decidable even when using $\IQ$ as an oracle set.
\end{theorem}
This result is, in spite of the
notational resemblance to $\IQ\subsetneq\IA$,
by no means obvious.
\subsection{Deciding $\IQ$ in $\IR$ by Means of an $\IA$--Oracle}
\label{secPos}
In this section, we prove

\begin{lemma}\label{lemma:Q-auf-A-reduzierbar}
~$\IQ\reduceq\IA$.
\end{lemma}

\begin{proof}
Consider some input $x\in\IR$.
By querying the $\IA$-oracle, identify and rule out the case that
$x$ is not in $\IA$ (and hence not in $\IQ$ either).
So it remains to distinguish $x\in\IQ$ from $x\in\IA\setminus\IQ$.
To this end, calculate $d:=\deg(x)$ according to Lemma~\ref{lemDeg}
below and test whether $d=1$ ~($x\in\IQ$)~ or ~$d\geq2$ ~($x\not\in\IQ$).
\qed
\end{proof}

\smallskip
\noindent
Recall that the \emph{degree} of an algebraic $a \in \IR$
is defined to be 
\[\deg(a)=\dim_{\IQ} \IQ(a)=[\IQ(a):\IQ],\] 
that is,
the dimension of the rational extension field generated by $a$.
It is well known, for example in
\mycite{Proposition~V.\textsection 1.2}{Algebra},
that finite field extensions 
$M \subset K \subset L$ satisfy
\begin{equation} \label{eqGradmult}
[L:M] \,=\, [L:K] \,\cdot\, [K:M] \enspace .
\end{equation}
A non-algebraic number is \emph{transcendental},
the set of which we shall denote by $\IT$.

\begin{lemma} \label{lemDeg}
The function 
$\deg:\IA\to\IN, \ a\mapsto\deg(a)$ 
is \BCSS--computable.
\end{lemma}
We point out that the restriction of $\deg$ to
algebraic numbers is essential here; in other words:
While for reasons of mathematical convenience
one can \emph{define}
$\deg(x):=\infty$ for transcendental $x$,
a \BCSS{} machine cannot \emph{compute} it.
\proof
Exploit that an alternative yet equivalent definition for $\deg(a)$
is given by the degree of a minimal polynomial of $a$,
that is, of an irreducible $p\in\IQ[X]$ of positive degree
with $p(a)=0$
\mycite{Proposition~V.\textsection 1.4}{Algebra}.
Moreover, $p$ can be chosen from $\IZ[X]$ with
content (i.e., the gcd of its coefficients equal to)
1. In this case, $p$ is irreducible in $\IQ[X]$
iff irreducible in $\IZ[X]$:
\textsf{Gauss' Lemma} \mycite{Theorem~IV.\textsection 2.3}{Algebra}.

Therefore we enumerate all non-constant $p\in\IZ[X]$ 
of content 1 and,
for each one, plug in $a$ to test whether $p(a)=0$.
If so, check $p$ for irreducibility ---
a property in classical \NP{} by virtue of \cite{Cantor}
and thus \BCSS--decidable.
If this test succeeds as well, return $\deg(p)$ and terminate;
otherwise continue with the next $p$.
\qed

\begin{remark}
An elementary decision procedure for irreducibility in $\IZ[X]$
proceeds 
--- although not within nondeterministic polynomial time ---
as follows:

Given $p\in\IZ[X]$ of degree $n-1>0$ and content 1, 
choose some $n$
arbitrary distinct arguments $x_1,\ldots,x_n\in\IZ$
and multi-evaluate $y_i:=p(x_i)$.
Observe that, if $q\in\IZ[X]$ is a non-trivial divisor of $p$,
then $z_i:=q(x_i)$ divides $y_i$ for each $i=1,\ldots,n$.
This suggests to go through all (finitely many)
choices for $(z_1,\ldots,z_n)\in\IZ^n$ with $z_i\mid y_i$,
to calculate the interpolation polynomial $q\in\IQ[X]$
to data $(x_i,z_i)$
and check whether its coefficients are integral and
$q$ divides $p$.
\end{remark}
\subsection{Undecidability of $\IA$ in $\IR$ with Support of a $\IQ$--Oracle}
\label{secNeg}
In this section, we prove ~$\IA\notreduceq\IQ$.

The undecidability of $\IA$ \emph{without} further oracle assistance
follows similarly to that of $\IQ$ from a continuity argument,
observing that each, $\IA$ and $\IQ$ as well as their complements,
are dense in $\IR$. In fact, algebraic numbers remain dense even
when restricting to arbitrary high degree:
\begin{lemma} \label{lemThree}
Let $x\in\IR$, $\varepsilon>0$, and $N\in\IN$. \\
Then, there exists an algebraic real $a$ of $\deg(a)=N$
with $|x-a|<\varepsilon$.
\end{lemma}
\begin{proof}
Take some arbitrary algebraic real $b$
of degree $N$, such as $b:=2^{1/N}$.
Since $\IQ$ is dense in $\IR\ni y:=x-b$,
there exists some rational $r\in\IQ$ with
$|r-y|<\varepsilon$.
Then $a:=r+b$ has the desired property.
\qed
\end{proof}
\smallskip
Of course, total discontinuity does not prevent a problem to be
\BCSS--decidable under the support of a $\IQ$--oracle any more as,
for example, $\IQ$ now is decidable. More precisely a putative
algorithm might try distinguishing algebraic from transcendental
reals by mapping a given $x$ through some rational function
$f\in\IR(X)$,
then querying the oracle whether the value $f(x)$ is rational
or not, and proceeding adaptively depending on the answer.

The following observation basically says that in
any sensible such approach, for transcendental $x$,
$f(x)$ will be irrational rather than rational.
\begin{lemma} \label{lemTwo}
Let $f:\dom(f)\subseteq\IR\to\IR$ be analytic
and non-constant, $T\subseteq\dom(f)$ uncountable.
Then, $f$ maps some $x\in T$ to a transcendental value,
that is, $f(x)\not\in\IA$.
\end{lemma}
\begin{proof}
Consider an arbitrary $y\in\IA$;
by uniqueness of analytic functions 
\mycite{Theorem~10.18}{Rudin},
$f$ can map
at most countably many different $x\in\dom(f)$
to that single value $y$.
Hence, if $f(x)\in\IA$ for all $x\in T$,
$f^{-1}(\IA)=\bigcup_{y\in\IA} f^{-1}(\{y\})$
is a countable union of countable sets and
thus countable, too --- contradicting the
prerequisite that $T\subseteq f^{-1}(\IA)$ is uncountable.
\qed
\end{proof}
So it remains the case of an algorithm trying to map
algebraic $x$ to rationals $f(x)$ and transcendental
$x$ to irrational $f(x)$.
The final ingredient formalizes the intuition that
this approach cannot distinguish transcendentals
from algebraic numbers of sufficiently high degree:
\begin{proposition} \label{lemOne}
Let $f\in\IR(X)$, $f=p/q$ with polynomials $p,q$
of $\deg(p)<n$, $\deg(p)<m$.
Let $a_1,\ldots,a_{n+m}\in\dom(f)$ be distinct
real algebraic numbers with $f(a_1),\ldots,f(a_{n+m})\in\IQ$.
\begin{enumerate}
\item[a)]
  There are co-prime polynomials $\tilde p,\tilde q$
  of $\deg(\tilde p)<n$, $\deg(\tilde q)<m$ with coefficients in
  the algebraic field extension $\IQ(a_1,\ldots,a_{n+m})$
  such that, for all $x\in\dom(f)=\{x:q(x)\not=0\}\subseteq\IR$,
  it holds $f(x)=\tilde f(x):=\tilde p(x)/\tilde q(x)$.
\item[b)] Let $d:=\max_i \deg(a_i)$. Then
  $f(x)\not\in\IQ$ for all transcendental
  $x\in\dom(f)$ as well as for all $x\in\IA$ of
  $\deg(x)>D:=d^{n+m}\cdot \max\{n-1,m-1\}$.
\end{enumerate}
\end{proposition}
Notice that $p$ and $q$ themselves in general do not
satisfy claim~a); e.g. $p=\pi\cdot\tilde p$ and $q=\pi\cdot\tilde q$.
\proof
a) \ Without loss of generality take $p$ and $q$ to be co-prime.
Let $y_i:=f(a_i)$.
The idea is to solve the rational interpolation problem
for $(a_i,y_i)$. Already knowing that is \emph{has} a
solution (namely $p,q$) avoids many of the difficulties
discussed in \cite{Rational}.

More precisely, observe that the
coefficients $p_0,\ldots,p_{n-1},q_0,\ldots,q_{m-1}\in\IR$ of
$p$ and $q$ satisfy the homogeneous
$(n+m)\times(n+m)$-size system of linear equations
$$ 
  \left( \begin{array}{ccccc|ccccc}
  1 & a_1 & a_1^2 & \ldots & a_1^{n-1}
  & -y_1 & -y_1 a_1 & \ldots & - y_1 a_1^{m-1} \\
  1 & a_2 & a_2^2 & \ldots & a_2^{n-1}
  & -y_2 & -y_2 a_2 & \ldots & - y_2 a_2^{m-1} \\
  1 & a_3 & a_3^2 & \ldots & a_3^{n-1}
  & -y_3 & -y_3 a_3 & \ldots & - y_3 a_3^{m-1} \\
  \vdots & \vdots & \vdots & \ddots & \vdots &
  \vdots & \vdots & \ddots & \vdots \\
  \end{array} \right)
   \quad\cdot\quad \left(\begin{array}{l} p_0 \\ \vdots \\ p_{n-1} \\
   q_0 \\ \vdots \\ q_{m-1}
   \end{array}\right) \quad = \quad 0 \enspace .
$$ 
In particular, this system has
$(p_0,\ldots,q_{m-1})\in\IR^{n+m}$
as non-zero solution.

The coefficients of the matrix live in 
$\IQ(a_1,\ldots,a_{n+m})$. Therefore, \person{Gau\ss{}}ian
Elimination yields a (possibly different) non-zero solution
$(\bar p_0,\ldots,\bar q_{m-1})$, also with entries
in $\IQ(a_1,\ldots,a_{n+m})$.
Now apply the \person{Euclidean} Algorithm to
the thus obtained polynomials $\bar p,\bar q$
and calculate their greatest common divisor $\bar h$ which,
again, has coefficients in $\IQ(a_1,\ldots,a_{n+m})$.

Thus, $\tilde p:=\bar p/\bar h$ and $\tilde q:=\bar q/\bar h$
are co-prime polynomials over $\IQ(a_1,\ldots,a_{n+m})$ of
$\deg(\tilde p)<n$ and $\deg(\tilde q)<m$ such that
$\tilde p\cdot q$ coincides with $p\cdot\tilde q$ on
arguments $a_1,\ldots,a_{n+m}$. This implies the
latter polynomials of degree less than $n+m$ to
be identical: $\tilde p\cdot q=p\cdot\tilde q$.

It follows that $q$ divides both sides; and co-primality of $(p,q)$
in the factorial ring $\IR[X]$ requires that
$q$ divides $\tilde q$. Similarly, $\tilde q$ divides $q$,
yielding $\tilde q=\lambda q$ for some $\lambda\in\IR$.
Analogously, $\tilde p=\lambda p$ for the same $\lambda$.

b) \ Consider $x\in\IR$ with $y:=f(x)\in\IQ$
and suppose $x$ is algebraic of
$\deg(x)>d^{n+m}\cdot \max\{n-1,m-1\}$ or transcendental.
Being, by virtue of a), a zero of the polynomial
$\tilde p-y\cdot\tilde q$
with coefficients from $\IQ(a_1,\ldots,a_n)$,
$x$ lies in an algebraic extension of the latter field,
hence ruling out the case that it is transcendental.
More precisely, the degree of $x$
{over $\IQ(a_1,\ldots,a_n)$} is 
bounded by
$\deg(\tilde p-y\cdot\tilde q)$;
and $\deg(x)$, its degree over $\IQ$,
is at most 
$\deg(\tilde p-y\cdot\tilde q)\cdot\deg(a_1)\cdots\deg(a_{n+m})
\leq \max\{n-1,m-1\}\cdot d^{n+m}$
by Equation~(\ref{eqGradmult})
--- contradiction.
\qed


\smallskip
\noindent
We are finally in the position to prove
\setcounter{theorem}{3}
\begin{theorem}\label{Proposition:A-nicht-mit-Q-Orakel}
In the \BCSS{} model of real number computation it holds
~$\IQ\reducneq\IA$. In particular, transcendence is not
semi-decidable even when using $\IQ$ as an oracle set.
\end{theorem}
\setcounter{theorem}{10}
\begin{proof}
Suppose some \BCSS{} algorithm semi-decides $\IT$ in $\IR$ with oracle $\IQ$
according to Definition~\ref{defBCSS};
in other words, it proceeds by repeatedly evaluating
a given $x\in\IR$ at functions $f\in\IR(X)$
and continuing adaptively according to whether
$f(x)$ is positive/zero/negative  and  rational/irrational,
such as to terminate iff $x\in\IT$.

Consider this process unrolled into an (infinite yet countable)
\textsf{Decision Tree},
each internal node $u$ of which is labeled with an according
$f_u\in\IR(X)$ and has five successors according to the cases
\begin{itemize}
\item[\textbullet]
$0>f_u(x)\in\IQ$ \hfill
\item[\textbullet]
$0>f_u(x)\not\in\IQ$ \hfill
\item[\textbullet]
$0=f_u(x)$ \hfill
\item[\textbullet]
$0<f_u(x)\in\IQ$ \hfill
\item[\textbullet]
$0<f_u(x)\not\in\IQ$
\end{itemize}
with leafs corresponding to terminating computations,
that is, to $x\in\IT$. Observe that the sets $T_v$ of
$x\in\IT$ terminating in leaf $v$ give rise to a
partition of $\IT$. In fact, the  at most countably
many leafs --- as opposed to $\IT$ having cardinality of
the continuum --- require that $T_v$ is uncountable
for at least one $v$.

Consider the path leading from the root to that leaf.
W.l.o.g. it contains no branches of type ``$0=f_u(x)$''
nor of type ``$f_u(x)\in\IQ$'' that are answered ``yes'';
for if it does, then the uncountable set
$T_v$ of transcendentals $x$ passing through this
branch implies that $f_u$ is constant (Lemma~\ref{lemTwo})
and node $u$ thus is dispensable.
By possibly changing from $+f_u$ to $-f_u$, we may finally
suppose that every branch on the path to leaf $v$ is of
type $0<f_u(x)$.

Summarizing,
$T_v\not=\emptyset$ is the set of exactly those
$x\in\IR$ satisfying
$0<f_u(x)\not\in\IQ$
for the (finitely many) internal nodes $u$
on the path from the root to $v$;
in particular, $T_v\subseteq\dom(f_u)$.
Now take some $t\in T_v\subseteq\IR$.
Due to continuity of rational functions,
there exists $\varepsilon>0$
such that $f_u(x)>0$ on all nodes $u$ on that path
for any $x\in\IR$ satisfying $|x-t|<\varepsilon$.
In particular, $f_u(a)>0$ holds for infinitely many
algebraic numbers $a$ of unbounded degree
according to Lemma~\ref{lemThree}.
Since by presumption, none of them completes the
(terminating) computational path to leaf $v$,
they must branch off somewhere,
that is, satisfy $f_u(a)\in\IQ$ for some of the
finitely many nodes $u$.
However by Proposition~\ref{lemOne}b), each single $f_u$
can sort out only algebraics of degree up to some finite
$D=D(u)$ ---
a contradiction.
\qed
\end{proof}

\section{More Undecidable and Incomparable Real Degrees} 
\label{secExistence}

A further achievement of the works of \person{Friedberg}
and \person{Muchnik} was the existence of 
incomparable r.e. degrees below the Halting problem.
In this section,
we extend our above techniques to establish
in the real case such problems explicitly.

More precisely, we shall construct natural incomparable
subsets of $\IA.$ They are given as certain algebraic, infinite
extensions of $\Q$ obtained by means of adjunction of $n$-th roots
of a fixed prime.

For simplicity, we consider two incomparable problems only.
However the construction immediately generalizes
to an infinite number of incomparable
real r.e. Turing degrees.

\subsection{Some Auxiliary Results from Algebra}
\label{Algebra-Resultate}

Consider the following type of algebraic extensions:
\begin{definition} \label{d:Wurzeln}
For fields $\IQ\subseteq\IF\subseteq\IR$ and $0<r\in\IQ$, let
\[ \IF(\sqrt[*]{r}) \quad:=\quad
  \IF\big( \big\{ r^{\frac{1}{n}} : n\in\IN\big\}\big) \]
where the corresponding fractional powers are understood as
positive real numbers.
\end{definition}
Thus, $\IQ(\sqrt[*]{2})$ results from $\IQ$ by field adjunction
of all $n$-th roots of $2$, $n \in \IN.$ 
The ancient proof of $\sqrt{2}$'s irrationality immediately generalizes
to see that this is indeed an infinite extension.
By Lemma~\ref{l:Wurzeln}c) below, this extends from $\IQ$ to,
e.g., $\IQ(\sqrt[*]{3})$.
In combination with Lemma~\ref{l:Wurzeln}d), it generalizes 
Lemma~\ref{lemThree}.

\COMMENTED{
\begin{fact}\label{Satz:Lang}
Fix some subfield $K$ of $\IR$ and $n\in\IN$.
If $0<a \in K$ is no $p$-th power
of an element in $K$ for any prime divisor $p$ of $n$,
then the field extension
$K(\sqrt[n]{a})$ has degree $n$ over $K$.
\end{fact}
\begin{proof} See \cite{Algebra}, Chapter 6, Theorem 9.1;
and observe that negative $-a/4$ cannot be an even
power in $K\subseteq\IR$ anyway.
\end{proof}
}

\begin{lemma}\label{l:Wurzeln}
\begin{enumerate}
\item[a)]
  If ~$(\tfrac{r}{s})^{1/n}\in\IQ$ ~for
  ~$n\in\IN$ and coprime $r,s\in\IN$,
  then $r^{1/n},s^{1/n}\in\IN$.%
\item[b)]
  For $n_1,\ldots,n_k\in\IN$ and squarefree $t\in\IN$, 
  $\IF\big(\sqrt[n_1]{t},\ldots,\sqrt[n_k]{t}\big)=\IF\big(\sqrt[N]{t}\big)$
  where $N:=\lcm(n_1,\ldots,n_k)$ denotes the \emph{least common multiple}.
\item[c)]
  For distinct prime numbers $p_1,\ldots,p_d,p_{d+1}$ and $n\in\IN$, 
  it holds\footnote{We owe considerable gratitude to \person{Toma Albu}
for pointing us to \cite{Besicovitch}}
  \[ \big[\IQ\big(\sqrt[*]{p_1},\sqrt[*]{p_2},\ldots,\sqrt[*]{p_d},
		    \sqrt[*]{p_{d+1}}\big)
    \,:\,
    \IQ\big(\sqrt[*]{p_1},\sqrt[*]{p_2},\ldots,\sqrt[*]{p_d}\big)\big] 
    \quad = \quad \infty \enspace . \]
\item[d)]
  To any $n\in\IN$, $\epsilon>0$, and $x\in\IR$,
  there exists $y\in\IQ(\sqrt[*]{2})$ of degree at least 
  $n$ over $\IQ(\sqrt[*]{3})$ such that $|x-y| < \epsilon$.
\end{enumerate}
\end{lemma}
\begin{proof}
\begin{enumerate}
\item[a)]
  W.l.o.g. $n\geq 2$. Let $(\tfrac{r}{s})^{1/n}=\tfrac{a}{b}$
  with coprime $a,b\in\IN$. Then
  any prime divisor $p$ of $s$ divides $s\cdot a^n=r\cdot b^n$
  but not $r$ (by coprimality) and thus $b^n$. Hence
  even $p^n$ divides $r\cdot b^n=s\cdot a^n$, so
  $p^n\mid s$. This reveals that every prime factor
  $p$ of $s$ occurs in $s$ with multiplicity a multiple of $n$,
  i.e., $s^{1/n}\in\IN$; similarly for $r^{1/n}$.
\item[b)]
Recall that the following properties of $\lcm$:
\begin{gather*}
n_i\mid\lcm(n_1,\ldots,n_k)=\lcm\big(\lcm(n_1,\ldots,n_{k-1}),n_k\big)
\\ \text{and }\quad
\lcm(a,b)=ab/\gcd(a,b)=ab/(ra+st) \quad\text{with } r,s\in\IZ \enspace .
\end{gather*}
Therefore, each $t^{1/n_i}$ is a power of $t^{1/N}$ 
and thus in $\IF(\sqrt[N]{t})$; while, conversely,
$t^{1/\lcm(a,b)}=(t^{1/a})^s\cdot(t^{1/b})^r\in\IF(\sqrt[a]{t},\sqrt[b]{t})$
yields $t^{1/N}$ to belong to $\IF(\sqrt[n_1]{t},\ldots,\sqrt[n_k]{t})$
by induction on $k$. Hence we have indeed established $t^{1/N}$ as
a primitive element.
\item[c)]
  \person{Besicovitch} has been proven that
  \[ \big[ \IQ\big(\sqrt[N_1]{p_1},\sqrt[N_2]{p_2},\ldots,\sqrt[N_d]{p_d}\big) 
    \,:\, \IQ
     \big] \quad=\quad N_1\cdot N_2\cdots N_d \enspace ; \]
  cf. \mycite{Theorem~2}{Besicovitch}; see also \cite[bottom of p.2]{Albu}.
  Now combine with Claim~b) and Equation~(\ref{eqGradmult}).
\item[d)]
  By c), $b:=2^{1/n}$ has
  degree $n$ over $\IQ(\sqrt[*]{3})$; and so has
  $y:=b+r$ for any $r\in\IQ$. $\IQ$ being dense,
  take $r$ close to $x-b$.
\qed
\end{enumerate}
\end{proof}
\subsection{Construction of Incomparable Degrees}
\label{unvergleichbar}

The tools from the previous subsection allow to extend
our results to obtain

\begin{theorem}\label{Satz:unvergleichbar}
The sets 
$\IQ(\sqrt[*]{2})$ and $\IQ(\sqrt[*]{3})$ 
are recursively enumerable yet incomparable.
\end{theorem}
Its proof is based on the following immediate generalization
of Proposition~\ref{lemOne}.
\begin{proposition} \label{lemExtOne}
   Let $f\in\IR(X), f = \frac{p}{q}$ with polynomials
   $p,q$   
   of degree less than $n$ and $m$, respectively. Let
   $a_1,\ldots,a_{n+m}\in \IQ(\sqrt[*]{2})\cap\dom(f)$ be distinct with
   $f(a_i) \in \IQ(\sqrt[*]{3})$.  
\begin{enumerate}
\item[a)]
 There are co-prime polynomials $\tilde p,\tilde q$
  of $\deg(\tilde p)<n$, $\deg(\tilde q)<m$ with coefficients in
  the algebraic field extension $\IQ(\sqrt[*]{3};a_1,\ldots,a_{n+m})$
  such that, for all $x\in\dom(f)=\{x:q(x)\not=0\}\subseteq\IR$,
  it holds $f(x)=\tilde f(x):=\tilde p(x)/\tilde q(x)$.
\item[b)] Let $d:=\max_i \deg_{\IQ(\sqrt[*]{3})}(a_i)$.
Then $f(x) \not\in \IQ(\sqrt[*]{3})$
for all transcendental $x \in \dom(f)$ as well as for all
$x \in \IQ(\sqrt[*]{2})$ of 
$\deg_{\IQ(\sqrt[*]{3})}(x) > D:=d^{n+m}\cdot \max\{n-1,m-1\}$.
\end{enumerate}
\end{proposition}

\begin{proof}[of Theorem \ref{Satz:unvergleichbar}]
For semi-decidability observe that,
with $N=\{n_1,n_2,\ldots\}$ and
due to Lemma~\ref{l:Wurzeln}b)
and \mycite{Proposition~\textsection V.1.4}{Algebra},
\begin{multline*}
\IQ(\sqrt[*]{2}) \;=\; \big\{
  x\in\IQ \:\big|\: \exists k\in\IN \: 
  \exists a_0,\ldots,a_{K-1}\in\IQ \\
  \underbrace{\exists t\in\IR:\:
    x=a_0+t a_1+\ldots+t^{K-1}a_{K-1}\,\wedge\, t^{K}=2}_{
   =:\Phi(N;a_0,\ldots,a_{K-1};x)}\big\}
\end{multline*}
where $K:=n_1\cdots n_k= [\IQ(2^{1/K}):\IQ]$.
Now $\Phi$ is an $\NP_{\IR}$--formula and thus decidable
by eliminating quantification with respect to $t$;
see, e.g., \mycite{Section~2.4}{Basu}.

Consider a putative machine semi-deciding $\IR \setminus
\IQ(\sqrt[*]{2})$ by means of an $\IQ(\sqrt[*]{3})$-oracle.
Follow the proof of Theorem~\ref{Proposition:A-nicht-mit-Q-Orakel}
and apply Lemma~\ref{lemTwo}
to obtain in just the same way a leaf $v$ together with the related
path set $T_v \subseteq \IR \setminus \IQ(\sqrt[*]{2})$.
Since $T_v$ is uncountable it contains a transcendental
$x$ and in each neighborhood of $x$ by virtue of 
Lemma~\ref{l:Wurzeln}d) 
elements of $\IQ(\sqrt[*]{2})$ of arbitrarily high degree
over the field $\IQ(\sqrt[*]{3})$. Thus, 
applying Proposition \ref{lemExtOne} 
there exist elements
in $\IQ(\sqrt[*][2])$ that are branched along $v,$ contradicting
the assumption that the machine semi-decides $\IR \setminus \IQ(\sqrt[*]{2})$.

The converse claim ``$\IQ(\sqrt[*]{3})\notreduceq\IQ(\sqrt[*]{2})$''
follows similarly.
\qed
\end{proof}
The numbers 2 and 3 in the above proof can obviously
be replaced by any two distinct primes; that is,
the sets $\IQ(\sqrt[*]{p})$ and
$\IQ(\sqrt[*]{q})$ are incomparable
for any two $p,q\in\IP=\{2,3,5,7,11,13,17,\ldots\}$.
In particular, we have \emph{explicitly}
an infinite number of incomparable degrees.
Moreover the argument immediately extends to see that, for
$P,Q\subseteq\IP$,
\[\IQ(\{\sqrt[*]{p}:p\in P\}) \;\reduceq\;\IQ(\{\sqrt[*]{q}:q\in Q\})
\qquad\Longleftrightarrow\qquad
P\subseteq Q \enspace . \]
Since the collection of subsets with inclusion
is the prototype of a poset,
we have thus arrived at the following
\begin{scholiumf}\footnotetext{A scholium is ``\emph{a note
amplifying a proof or course of reasoning, as in mathematics}''.}
Every countable poset 
can be embedded into the recursively enumerable real Turing degrees.
\qed
\end{scholiumf}
This parallels classical results in discrete recursion theory;
see for instance
\cite[\textsc{Exercise~\textsection VII.2.2}(b)
and \textsc{Exercise~\textsection VIII.4.10}]{Soare}.

\subsection{Some Open Problems}
The previous arguments lead to some other problems
concerning the relation between some natural
subsets of $\IR$ that we consider to be interesting.

For  $d \in \IN$ let 
$\IA_d:=\{x\in\IA:\deg(x)\leq d\} \subset \IR$ denote the set of
algebraic numbers that have degree at most $d$ over $\IQ.$

\begin{myproblem}
Is it true that we have a strict chain
\[\IQ \;\reducneq\; \IA_2 \;\reducneq\; \IA_3 \;\reducneq\; \ldots 
\reducneq\; \IA \;\reducneq\; \IH \enspace ? \]
\end{myproblem}
We have defined $\IA_d$ to consist of numbers of degree
\emph{less or} equal to $d$ but point out that considering,
rather than $\IA_2=:\IA_{\mysmall{\leq}2}$,
the set $\IA_{\mysmall{=}2}:=\{x\in\IA:\deg(x)=2\}$ of numbers
of degree \emph{exactly} 2, in fact makes no difference:
\begin{lemma} It holds ~$\IA_{\mysmall{=}2} \,\equiv\, \IA_{\mysmall{\leq}2}$.
\end{lemma}
\begin{proof}
Based on oracle access to $\IA_{\mysmall{\leq}2}$, decide $\IA_{\mysmall{=}2}$
in $\IR$ as follows: Upon input of $x\in\IR$, query
$\IA_{\mysmall{\leq}2}$ to find out whether $\deg(x)\leq2$. If not,
reject; otherwise $x\in\IA$ and we may apply Lemma~\ref{lemDeg} 
to compute $\deg(x)$.

Conversely, given $\IA_{\mysmall{=}2}$ as an oracle, decide 
whether $x\in\IA_{\mysmall{\leq}2}$ by querying both $x$
and $y:=x+\sqrt{2}$. If at least
one of them belongs to $\IA_{\mysmall{=}2}$, then $x$ is
surely algebraic and thus applicable to Lemma~\ref{lemDeg}.
If $x,y\in\IR\setminus\IA_{\mysmall{=}2}$, we may reject immediately
because $\deg(x)<2$ would imply $x\in\IQ$ and thus 
$y=x+\sqrt{2}\in\IA_{\mysmall{=}2}$.  \qed
\end{proof}
But what about this question for general degrees $d \in \IN?$
\begin{myproblem}
Does it hold
~$\IA_{=d} \equiv \IA_{\leq d}$~
for all $d\geq 2$ ?
\end{myproblem}
Another interesting question has been kindly pointed out by a 
referee:%
\begin{myproblem}
Is there a \emph{countable} set Turing-equivalent
to the real halting problem $\IH ?$
\end{myproblem}
A disproof of the latter would, just by reasons
of cardinality, include and significantly strengthen
our result $\IH \notreduceq \IQ$ but not the stronger
claim $\IA \notreduceq\IQ$.

\section{The Linear \BCSS{} Model}\label{secLinear}

We have so far considered the full \BCSS{} model over the reals.
In the last ten years, its linearly restricted version
$(\IR,+,-,0,1,<)$ has received increasing interest
\cite{Koiran,CuckerKoiran,Michaux} due to its relation with
the classical (i.e., discrete) 
``$\mathcal{P}\overset{?}{=}\NP$'' question
\cite{Fournier}.
Here only additions, subtractions and
comparisons as well as the constants $0$ and $1$ 
are allowed but no multiplication $\times$ nor division $\div$.
Thus, all computed intermediate results
on inputs $x \in \IR$ have the form $ax+b$ 
for some $a,b \in \IZ.$
Analogously to the full model,
the Halting Problem for linear machines is undecidable
by a linear machine;
and Post's problem as well makes sense in the linear version.
In order to give an explicit solution to it, 
we once more consider the rationals $\IQ,$ but this time 
as the harder of two problems. 
The weaker undecidable one will be the following:

\begin{definition} Let 
~$\SQ:=\{q^2:q\in\IQ\}$~ 
denote the set of quadratic rationals.
\end{definition}
We shall show that $\SQ \reducneq \IQ$, where in this section
``$\reduceq$'' and all similar notions refer to reducibility 
in the linear model. 
We start with some easy observations. Both $\IQ$ and $\SQ$ are undecidable
in the linear model since this already holds in the full model. 
Both sets are semi-decidable:
For input $ x \in \IR$ enumerate all pairs $(r,s) \in \IZ \times \IN$
and check for each pair whether $ x \cdot s = r.$ Note that both the enumeration
and the `multiplication' ~$x \cdot s$ can be performed in $(\IR,+,-,0,1,<);$
similarly for semi-deciding $\SQ$  by enumerating all pairs $(r^2,s^2)$ 
based for instance on the recursion $(r+1)^2=r^2+r+r+1$.
Next, $\SQ \reduceq \IQ$: On input $x \in \IR$, first check $x\geq0$
and ask the $\IQ$-oracle whether $ x \in \IQ.$ If this is the case use the 
above enumeration to find
$(r,s) \in \IN^2$ with $xs=r$.
Then test whether some of the (finitely many) pairs ~$(\tilde r^2,\tilde s^2)\leq(r,s)$ 
satisfies $x\cdot\tilde s^2=\tilde r^2$ or not.

Note that in the full \BCSS{} model the converse 
relation $\IQ \reduceq \SQ$ is also valid: Having access to a $\SQ$--oracle
one can decide $\IQ$ by simply squaring the input $x \in \IR.$
The main result of this section reveals that this reduction
does not hold in the linear model:

\begin{theorem}\label{Postlinear}
In the linear \BCSS{} model, it is $\SQ \reducneq \IQ$.
\end{theorem}
The proof applies Lemmas~\ref{lemmaaffin} and \ref{lemmadicht}
which are in some sense linear counterparts to Proposition~\ref{lemOne}b)
and Lemma~\ref{lemThree}, respectively.

\begin{lemma}\label{lemmaaffin}
Let $P\subseteq\IP$ be a (finite or infinite) set of primes. Define 
\[ 
\QP \;:=\;
\big\{ \tfrac{r}{s} \;:\; r\in\IZ,\,s\in\IN\,\wedge\, 1=\gcd(r,s)
\,\wedge\, \sqrt{s}\not\in\IN 
\,\wedge \, \forall p\in\IP: (p\mid s \Rightarrow p\in P)
\big\} 
\]
as the set of rationals whose
denominator, in reduced form with respect to the
numerator, is no square and contains only prime factors from $P$.
This satisfies
\begin{itemize}
\item[a)] $\QP \cap \SQ = \emptyset$.
\item[b)] Let $a \in \IZ$ having no prime factors $P$ and $b\in\IZ$. \\
Then $x \in \QP$ implies $y:=a \cdot x + b \in \QP$.%
\end{itemize}
\end{lemma}

\begin{proof} a) is a special case of 
Lemma~\ref{l:Wurzeln}a).
For b) suppose  that $x:= \frac{r}{s} \in \QP$ 
with coprime $r,s$ and $a,b$ as in the statement. 
Then $y=\tfrac{ar + bs}{s}$ with $\gcd(ar+bs,s) =1$;
the latter holds because a putative prime factor $p$
of $\gcd(ar+bs,s)\mid s$ 
belongs to $P$ by definition
and thus does not divide $a$ nor $r$, contradiction.
In particular, the reduced denominator $s$ of $x$
is also that of $y$.
\qed
\end{proof}

\begin{lemma}\label{lemmadicht}
For each $p\in\IP$, the set
~$
\tilde\IQ_p
:=
\big\{{r}/{p^{2k+1}}\,:\,k\in\IN, \, r\in\IZ, \, p\nmid r\big\}
\subseteq\IQ
$~ 
is dense in $\IR$.
In particular, so is $\QP$ for any non-empty $P\subseteq\IP$.
\end{lemma}
\begin{proof} 
The (not necessarily reduced) \textsf{$p$-adic rationals}
~$\IQ_p:=\{{t}/{p^k}:r\in\IZ,k\in\IN\}$~
are obviously dense: To $x\in\IR$ and arbitrary
$k\in\IN$, let $t:=\lfloor x\cdot p^k\rfloor\in\IZ$.

Now to $y={t}/{p^k}\in\IQ_p$ take any $n\in\IN$ and let
$k:=\ell+n$, $r:=t\cdot p^{\ell+1+2n}+1$. Then
$p\nmid r$, so $z:={r}/{p^{2k+1}}$ belongs to
$\tilde\IQ_p$
; and $|z-y|=p^{-(2n+2\ell+1)}$ becomes arbitrarily
small in $n$. Hence $\tilde\IQ_p$ 
is dense in $\IQ_p$ and thus in turn in $\IR$ as well.

Finally, $\QP$ is a superset of $\tilde\IQ_p$ for $p\in P$.
\qed
\end{proof}

\begin{proof}[Theorem~\ref{Postlinear}]
As usual we take a potential linear $\SQ$-oracle machine $M$ 
semi-deciding $\IR\setminus\IQ$ and pick a certain input $z>0$ 
which this time suffices to be chosen as irrational.
Let $f_i:x\mapsto a_i\cdot x+b_i$ denote the finitely
many test-functions evaluated on $z$ by $M$ before arrival
in a leaf, $a_i,b_i\in\IZ$, $1\leq i\leq I$.
Take $P\subseteq\IP$ such that $\IP\setminus P$
contains all (finitely many) prime factors 
of these coefficients $a_i$ and $b_i$.
Since $z$ is irrational, so is $f_i(z)\not\in\SQ$ and 
in particular $f_i(z)\not=0$ (w.l.o.g. $>0$)
for all $i$; hence it holds $f_i(x)>0$ for all
$i$ and all $x$ in some non-empty neighborhood of $z$.
By Lemma~\ref{lemmadicht} we can furthermore require $x\in\QP\subseteq\IQ$; 
by Lemma~\ref{lemmaaffin} for this $x$
all oracle queries ``$f(x)\overset{?}{\in}\SQ$'' 
are answered negatively.
In other words, $M$ branches
$x$ along the very same path as $z$
and eventually ends up in a leaf,
contradicting that $M$ terminates only
for $x\not\in\IQ$.
\qed
\end{proof}

\begin{myproblem}
In the linear setting,
does $\IQ$ have the same degree of undecidability as the Halting Problem?
\end{myproblem}

\section{Conclusion} \label{secConclusion}
We have shown that oracle access to the set of rational numbers
$\IQ$ gives a \BCSS{} machine additional power but still prevents it
from solving the real Halting Problem $\IH$ (of \BCSS{} machines).
In addition we have explicitly specified an
uncountable number of incomparable recursively enumerable 
degrees in the real number setting.
This involved arguments from topology as well as from
abstract algebra; e.g., transcendence, irreducible polynomials,
and finite field extensions play a major role.
In the linear setting, a similar result was obtained
using number theory; e.g., irrationality, primes,
and integral lattices.

Our proofs generally
do not rely on the ordering available over the real numbers.
Thus with small corrections (for example a slightly changed
definition of the 
characteristic path in a potential decision tree)
it also works over the complex numbers yielding
the corresponding results.

We close with some remarks concerning hypercomputation.
Since there is no commonly accepted definition of what hypercomputation 
should be our remarks, however, are a bit speculative.
Regarding attempts to physically realize hypercomputation over
the reals our results indicate that it seems advisable (since provably easier)
to construct a device capable of solving $\IQ$ rather than $\IH$.
Such an approach may, in contrast to discrete hypercomputation,
benefit from the explicit knowledge of this degree.

One might object that, since ~`\emph{Natura non facit saltus}'~
according to \person{Leibniz}, the discontinuity inherent in
deciding $\IQ$ in $\IR$ (i.e., of distinguishing fractions
from general reals) makes an according devise physically
impossible. However we point out that for example the
Fractional Quantum Hall Effect (Nobel Prize Physics 1998)
shows that nature does exhibit exactly
this kind of discontinuous behaviour.


\end{document}